\documentclass{article}
\usepackage[utf8]{inputenc}
\usepackage{threeparttable}
\usepackage{mathtools}
\usepackage{bm}
\usepackage{graphicx}
\usepackage{tabularx}
\usepackage{textcase}
\usepackage{dashrule}
\usepackage{ragged2e}
\usepackage{authblk}
\usepackage{soul}
\usepackage{xpatch}
\usepackage{afterpage}
\usepackage{float}
\usepackage{amsmath}
\usepackage{mathabx}
\usepackage{bbm}
\usepackage[table]{xcolor}
\usepackage{xcolor}
\usepackage{caption}
\usepackage{placeins}

\newcommand{\bgamma}{\mbox{\boldmath $\gamma$}}

\newcommand{\bbeta}{\mbox{\boldmath $\beta$}}

\newcommand{\btheta}{\mbox{\boldmath $\theta$}}
\newcommand{\balpha}{\mbox{\boldmath $\alpha$}}

\newcommand{\bfomega}{\mbox{\boldmath $\omega$}}

\AtBeginEnvironment{table}{%
  \rowcolors{2}{gray!25}{white}
}

\usepackage{threeparttable}

\usepackage[super]{natbib}
\date{}
\title{How much of the past matters?\\ Using dynamic survival models for the  monitoring of potassium in heart failure patients using electronic health records}

\author{Caterina Gregorio\textsuperscript{1,2}, Giulia Barbati \textsuperscript{2}, Arjuna Scagnetto \textsuperscript{3}, Andrea Di Lenarda \textsuperscript{3}, Francesca Ieva \textsuperscript{1,4}\\
\textsuperscript{1} MOX - Modelling and Scientific Computing, Department of Mathematics Politecnico di Milan,
Piazza Leonardo Da Vinci 32, Milan 20133, Italy,\\
\textsuperscript{2} Biostatistics Unit, Department of Medical Sciences, University of Trieste, Via Valerio 4$\backslash$1, Trieste 34100, Italy,\\
\textsuperscript{3} Territorial Specialistic Department, University Hospital and Health Services of Trieste, Italy,\\
\textsuperscript{4} HDS, Health Data Science center, Human Technopole,  Viale Rita Levi Montalcini 1,  Milan 20157, Italy.}

\begin{document}

\maketitle

\begin{abstract}
 
 Statistical methods to study the association between a longitudinal biomarker and the risk of death are very relevant for the long-term care of subjects affected by chronic illnesses, such as potassium in heart failure patients. Particularly in the presence of comorbidities or pharmacological treatments, sudden crises can cause potassium to undergo very abrupt yet transient changes.  In the context of the monitoring of potassium, there is a need for a dynamic model that can be used in clinical practice to assess the risk of death related to an observed patient's potassium trajectory. We considered different dynamic survival approaches, starting from the simple approach considering the most recent measurement,  to the joint model. We then propose a novel method based on wavelet filtering and landmarking to retrieve the prognostic role of past short-term potassium shifts. We argue that while taking into account past information is important, not all past information is equally informative.  State-of-the-art dynamic survival models are prone to give more importance to the mean long-term value of potassium. However, our findings suggest that it is essential to take into account also recent potassium instability to capture all the relevant prognostic information. The data used comes from over 2000 subjects, with a total of over 80 000 repeated potassium measurements collected through Administrative Health Records and Outpatient and Inpatient Clinic E-charts. A novel dynamic survival approach is proposed in this work for the monitoring of potassium in heart failure. The proposed wavelet landmark method shows promising results revealing the prognostic role of past short-term changes, according to their different duration, and achieving higher performances in predicting the survival probability of individuals. 
\end{abstract}

\section{Introduction}

In recent years, medical research has increasingly relied on the integration of administrative and electronic health recording systems, providing vast amounts of longitudinal data from real-world contexts. Such data are crucial in extracting real-world evidence that can help manage complex conditions, such as heart failure.\\ Heart failure is a consequence of many cardiovascular diseases. However,  despite improvements in treatments, mortality, and hospitalization rates remain high in heart failure disease.  Monitoring patients' disease progression and status over time using easily measurable biomarkers, such as potassium, is crucial for making medical decisions regarding treatments. Potassium alterations are common in heart failure patients due to the disease itself, pharmacological treatment, and comorbidities. Dyskalemia, or potassium disorders, can lead to life-threatening conditions, with both low (hypokalemia) and high (hyperkalaemia) levels being dangerous. Clinical guidelines set the normal range of serum potassium between 3.5–5.0 mmol/L in the general population. However, recent studies have raised serious concerns about the validity of this range in patients affected by heart failure \cite{Cooper2020,Ferreira2020}. This work has been motivated by the need to further study the role of potassium dynamics and to provide a tool for the monitoring of potassium in the health care of heart failure patients. However, potassium in heart failure patients is characterized by sudden but typically transient changes, making it challenging to study its temporal evolution. Since the final goal of medical decisions is the minimization of the risk of adverse events such as hospitalizations or death, dynamic survival models can be used to build prognostic clinical models to study the association between time-to-event outcomes and biomarkers over time. 
In this article, we consider different approaches to dynamic survival with potassium as the biomarker of interest using electronic health records data. This class of models allows taking into account that the value of biomarker changes over time and with that, the risk of adverse events needs to be updated. The most simple approach to dynamic predictions is the standard landmarking \cite{VanHouwelingen2011}. This approach resembles what is most typically done in clinical practice. Only the most recent measurement of the biomarker is taken into account. Moreover, the assessment of such values is done mainly through cut-offs which distinguish ``normal" values from ``abnormal" ones. The major drawback of this method is that past information given by repeated measurements is either not considered in the medical assessment or used qualitatively. On the other hand, it has been shown that exploiting all available information and quantitatively assessing the risk of an individual biomarker trajectory observed up to a time can be much more informative.  In the last few years, most of the literature involving dynamic survival models has been based on joint models \cite{Rizopoulos2012}. Joint models are the current gold standard in this setting, as they allow for past history and measurement error to be taken into account. In this class of models, the longitudinal biomarker process is modelled through linear mixed effects models, which allow considering the subjects' specific trajectories of the biomarker through the inclusion of random effects into the model. These latent variables are used to model the effect of unobserved variables that are responsible for subjects' deviation from the overall mean trajectory specified through the fixed effects \cite{TsiasisAnastasiosA.2004}.  Much of the research in this context has been devoted to studying different specifications of joint models \cite{Papageorgiou2019} and comparing them with standard or more advanced approaches based on landmarking, such as the landmarking coupled with linear mixed effect models \cite{Rizopoulos2017,Ferrer2019}. Overall, joint models have proved to successfully answer many research questions involving longitudinal biomarker data and survival outcomes in cardiovascular settings \cite{VanVark2017,DelaEspriella2020,Andrinopoulou2015,Kalogeropoulos2017}.
This work aims to explore the advantages and limitations of current statistical methods used to analyse the relationship between potassium dynamics and time-to-event outcomes.  In addition, a new approach using wavelets and landmarking is proposed to better understand the role of how past variations in potassium levels may affect mortality risk depending on the time window over which they occurred. Wavelet filters are common methods in time series analysis \cite{Nason1999} and signal processing \cite{Unser1996} to extract relevant features from data and remove noise. One of their characteristics is their ability to identify changes that occur at different frequencies. We believe that this can also be an insightful feature in the context of clinical monitoring of biomarkers such as potassium.  In Section 2 we describe the data used in this article. Section 3 describes the statistical methods. First, the concept of dynamic predictions is presented, and then the different methods considered in this article to obtain dynamic predictions are introduced: the landmark Last Observation Carried Forward (LOCF) model (Section 3.2), the landmark mixed survival model (Section 3.3), the Joint model for longitudinal and survival data (Section 3.4) and finally the novel landmark wavelet model (Section 3.5).  A simulation study was performed to evaluate the appropriateness of applying the proposed wavelet method to longitudinal potassium data (Section 4). The results of the real-world data application are given in Section 5. In Section 5.1 a general description of the cohort is given, while in Section 5.2 potassium repeated measurements data is described. The results of the different dynamic survival models are shown in Section 5.3. The performance and comparison of the different approaches in terms of goodness-of and dynamic predictive accuracy in the application under study is reported in Section 5.4. Finally, the discussion and conclusions are in Section 6.  All analyses were performed in R and the code is available at https://github.com/caterinagregorio/DynSurvPotassiumMonitoringHF.

\section{Study design and Data}

 Data was obtained by the interrogation of the administrative regional health data of Friuli Venezia Giulia Region in the Northern part of Italy, integrated with data derived from the Outpatient and Inpatient Clinic E-chart (Cardionet \textregistered). This integrated database constitutes the Trieste Observatory of Cardiovascular Diseases. Specifically, this was a cohort observational, non-interventional study involving patients living in the Trieste area who had a Heart Failure diagnosis between January 2009 and December 2020, had at least one cardiological evaluation, and two potassium measurements, and were observed for at least one year.  For the identification of the HF patients, the following steps were followed. First, a search in the electronic medical records, using appropriate keywords (heart failure, chronic Heart Failure, Systolic heart failure, diastolic heart failure) to select patients with HF-related clinical findings. In order to avoid any diagnostic underestimation, data from the medical E-chart were combined with the discharge codes of any previous hospital access (based on the standard nomenclature of the ICD-9 CM) and/or interventional procedures for HF patients (i.e. ICD implantation). Subsequently, prospective cases were manually reviewed by clinicians, to validate the diagnosis of HF using the criteria established in 2016 by the European Cardiology Society. The cohort was followed from the index date, defined as the date of the first ambulatory visit with a potassium measure available, until the time of death or the end of the follow-up (administrative study closure date, fixed at 31 December 2020).  The database has been previously described in the literature \cite{Iorio2019}. For this specific study, demographic, clinical and instrumental variables at the index date together with all repeated blood tests containing the potassium measurements have been considered.
 
\section{Methods}

Let $\mathcal{D}_n=\{Z_i,\delta_i,\textbf{y}_i; i=1,...,n\}$ denote a sample from the target population where $T_i$ and $C_i$ the individual's death and censoring time respectively. We assume that $C_i$ is non-informative with respect to the potassium process and death time. For each patient, we only observe the couple $Z_i=min(T_i,C_i)$, $\delta_i=\mathbbm{1}(T_i\le C_i)$. Moreover, for each patient $i \in \{1,...,n\}$, we let $\textbf{y}_i$ be the vector of longitudinal potassium measurements and $y_{ij}$ a  single measurement observed for the subject $i$ at time $t_{ij}$, $j=1,...,m_i$.

\subsection{Dynamic predictions}

The main objective is to provide a tool to extract from potassium longitudinal data observed up to a time, e.g. in correspondence of a clinical visit of patients affected by heart failure, a quantitative assessment of patients' future risk of death. Using dynamic survival models, it is possible to derive individualized dynamic predictions of survival. We let $k$ be the subject that has provided a set of longitudinal potassium measurements up to time $t$, $\mathcal{Y}_k(t)=\{y_k(t_{kj}); 0 \le t_{kj}\le t, j=1,...,m_k\}$. The individual dynamic prediction is defined for a specified time horizon $w>t$ as the probability that the subject $k$ will survive, at least up to $w$:

\begin{equation}
    \pi_k(w|t)=Pr(T_k\ge w|T_k>t,\mathcal{Y}_k(t),\mathcal{D}_n)
\end{equation}

Note that this is a conditional subject-specific prediction since the subject has survived to time $t$. The prediction is called dynamic in time because we can update it to get $\pi_k(w|t')$ as new information is collected at $t'>t$. 
In the following Sections, we describe different methods to obtain dynamic predictions of survival based on past potassium measurements.

\subsection{Landmark LOCF}

The standard conditional approach is one more similar to what is currently done in clinical practice. We can set a grid of time points, called landmarks, in which information on potassium can be updated. At each landmark point, we can study the association between potassium and time of death by estimating a Cox proportional hazard model using as a covariate the last potassium measurement collected before the landmark point. The dataset used for the estimation contains only patients still under observation at the landmark time. Moreover, only the events up to the fixed time horizon $w$ are retained, while patients experiencing the event after the horizon are censored at the horizon time. Let $h=\{h_1, ... ,t,..., h_l\}$ the landmark times.
To gain efficiency and interpretability, instead of estimating different models for each landmark time, we can specify a so-called ``landmark supermodel" \cite{VanHouwelingen2011}:

\begin{equation*}
 h_i(l|\textbf{z}_i, y_{ih}, h,w)=h_{h,0}(t) \exp\{\textbf{z}_i\bbeta+f(y_{ih})\gamma \}, h\le l\le h+w
\end{equation*}

where $\textbf{z}_i$ is the vector of fixed covariate, $f(y_{ih})$ is a generic transformation of the last potassium assessment before time $h$ and $\bbeta$ and $\bgamma$ are the respective coefficients. Specifically, two different models with two alternative transformations of the last potassium measurements are considered. The first (LOCF1) considers the last observed measurement of potassium as a continuous value. Since both low and high values of potassium are believed to be dangerous, the last potassium measurement is allowed to have a non-linear effect on the hazard through a cubic B-spline with 4 degrees of freedom (internal knots were placed at 25th, 50th and 75th percentiles of the distribution). In the second model (LOCF2), from the last measurement of potassium, a categorical variable is derived using the cut-offs currently employed in the clinical practice. Specifically, potassium is considered  ``high" if greater than 5 mmol/L, ``low" if below 3.5 mmol/L and, in the normal range otherwise. 
The dataset used for the estimation is the one made by stacking all the datasets used for the separate landmark models. It is important to note that an event can appear in the ``stacked" dataset more than once if it falls in the $h+w$ interval for more than a landmark point $h$. This is the reason why the robust sandwich estimator for the standard errors is used.

Once the model has been estimated, an estimate of $\pi_k(w|t)$  for a subject $k$ is obtained as:

\begin{equation*}
    \hat{\pi}_k^{LOCF}(w|t)=exp[-\hat{H}_{t,0}(w)\exp\{\textbf{z}_k\hat{\bbeta}+f(y_{kt})\hat{\gamma} \}]
\end{equation*}

where $\hat{H}_{t,0}(u)$ is the cumulative baseline hazard obtained using the Breslow estimator.

\subsection{Mixed Landmark model}

The LOCF landmark approach is relatively simple. However, it does not take into account any past information besides the last measurement or the fact that potassium, as any biomarker, is measured with error. To overcome these limitations, an extension to the previous model consists of modelling the individual potassium trajectories using a linear mixed effect model:

 \begin{equation}
    y_{it}=m_{it}+\epsilon_i(t)=(\balpha+\textbf{a}_{i})f(t)+\tilde{\textbf{x}}_i\tilde{\balpha}+\epsilon_i(t)
    \label{eqmixed}
\end{equation}

where $\epsilon_i(t)$ is the Gaussian measurement error term with mean 0 and variance $\sigma^2_{\epsilon}$, $f(\cdot)$ is a non-linear function of time, $\balpha$ and $\tilde{\balpha}$ are fixed-effect coefficients for time and baseline covariates respectively and $\textbf{a}\sim N(\textbf{0},\textbf{D}) $ are the random effects coefficients.  A flexible dependence between the overall mean trajectory and time is defined through a natural cubic spline with $d$ degrees of freedom (denoted in the following with $ns(\cdot)$). Subject-specific changes in the potassium trajectory can be determined by disease progression as well as drug intake, hospitalizations, and worsening of renal function. These are unobserved longitudinal processes, and their effect on potassium can be taken into account using time-related random effects. For the random effects, we consider a spline transform of time with the same degrees of freedom used for the fixed effects, to allow for the estimated individual potassium trajectories to be as flexible as possible.

Now, in the landmark model, we assume that the risk of death depends on a transformation of the estimated value of mean subject-specific trajectory at each of the landmark times\cite{Rizopoulos2017}:

\begin{equation}
 h_i(l|\textbf{z}_i,y_{i0},...,y_{ih}, h,w)=h_{h,0}(t) \exp\{\textbf{z}_i\bbeta+f(\hat{m}_{ih})\gamma_1 \}, h\le l\le h+w
 \label{landmixed}
\end{equation}

where $\hat{m}_{ih}$ can be predicted at the different landmark times by deriving the empirical Bayes estimates of the random effects and the estimated fixed effects from the linear mixed model in \ref{eqmixed}. As in the previous model, a cubic spline was used, as $f(\cdot)$ to allow $\hat{m}_{ih}$ to have a non-linear effect on the hazard of death.

In the same way as before, we can derive $\hat{m}_{kt}$ at time $t$ for the subject $k$ for whom we want to calculate the dynamic prediction based on measurements collected up to $t$. Therefore, similarly to the LOCF landmark approach, the dynamic predictions are obtained as:

\begin{equation}
    \hat{\pi}^{MIXED}_k(w|t)=exp[-\hat{H}_{t,0}(w)\exp\{\textbf{z}_k\hat{\bbeta}+f(\hat{m}_{kt})\hat{\gamma}_1 \}]
\end{equation}

\subsection{Joint Model}

A different approach to obtaining dynamic prediction is the joint model, in which the joint distribution of the longitudinal potassium outcome and the time-to-event outcome is specified.

Again, to describe the longitudinal potassium outcome, a linear mixed-effect model similar to the one used for mixed landmarking can be employed. For the survival process, we assume now that the risk of the event depends on the current value of the ``true" mean individual longitudinal trajectory of the potassium. The joint model is therefore specified as follows:

\begin{equation}
    y_{it}=m_{it}+\epsilon_i(t)=(\balpha+\textbf{a}_{i})f(t)+\tilde{\textbf{x}}_i\tilde{\balpha}+\epsilon_i(t)
\end{equation}

\begin{equation}
    h_i(t|\mathcal{H}_i(t),\textbf{z}_i)=h_0(t)\exp\{\textbf{z}_i\bbeta+f(m_{it})\gamma_1 \},t>0
\end{equation}

where $\mathcal{H}_i(t)={m_{is},0\le s\le t}$ denotes the history of the true unobserved mean longitudinal process up to time $t$. Because of computational advantages, the model is typically estimated under a Bayesian framework using Markov chain Monte Carlo (MCMC).

In this case, an estimate of the dynamic prediction can be derived from a Monte Carlo simulation scheme based on this formula \cite{Rizopoulos2012}:

\begin{equation}
    \pi^{JM}_k(w|t)=\int Pr (T\ge w| T>t, \mathcal{Y}_k(t),\bfomega )p(\bfomega|\mathcal{D}_n)d\bfomega
\end{equation}

where $\bfomega$ is the vector of the model coefficient. The first term is derived from the conditional independence assumption of the joint model which states that given the random effects, both longitudinal and event time processes are assumed independent, and the longitudinal responses of each subject are assumed independent.

\subsection{Mixed-Wavelet Landmark model}

The approaches that use the linear mixed effects model to extract information from potassium measurements place a lot of emphasis on the current value of the biomarker, which results from the slow and stable changes. They tend to treat transient changes as measurement errors to be forgotten. Therefore, both joint and mixed landmark models implicitly assume that past short-term changes in potassium don't affect the risk of adverse events. However, in patients with heart failure, past changes in potassium that are not reflected in the current value may be important to consider when monitoring potassium.  
To assess whether short-term changes play a role in shaping the risk of death, we propose a novel method based on landmarking and the wavelet Morlet filter \cite{Carmona1998}.  In general, Morlet wavelets are continuous, complex-valued functions used to smooth non-stationary time-series data, and they allow for distinguishing the frequencies at which oscillations occur. The Morlet wavelet transform is characterized by a ``mother wavelet": $\Psi(t)=\pi^{-1/4}e^{6it}e^{-t^2/2}$.
In the application to dynamic survival models and potassium monitoring, we propose the use of the Morlet filter to extract from the repeated measurements' data, the times at which relevant changes happen.
To achieve this, the data needs to be de-trended so the Morlet transform is applied to the residuals of the linear mixed effect model specified in Equation \ref{eqmixed}, denoted by $\tilde{y}_{it}$. The set of frequencies which will determine the duration of the short-term changes that are allowed to be captured by the filtering also needs to be chosen. Here, we define equivalently their reciprocal, the periods (or duration) $p$.
The short-term oscillation in potassium at time $t$ and of duration within $p_1 \le p \le p_2$ is obtained using the  Morlet wavelet transform as \cite{Roesch2018}:

\begin{equation*}
    \hat{s}^{p_1-p_2}_i(\tau)=\frac{df dt^{1/2}}{0.776 \Phi(0)}\sum_{p=p_1,...,p_2} \hat{\theta}^{\tau,p}_i\frac{Re[Wave_i(\tau,p)]}{(1/p)^{1/2}}
\end{equation*}

where $Wave_i(\tau, p)=\sum_t \tilde{y}_{i}(t) \frac{1}{1/\sqrt{p}}\Psi^* \big(\frac{t-\tau}{1/p}\big)$,  $\Psi^*$ indicates the complex conjugate of $\Psi$. $\tau$ determines the position of a specific daughter wavelet in time, and $df$ and $dt$ are the frequency and time resolutions chosen. Finally, $Re(\cdot)$ denotes the real part. The filtering itself is carried out by the term $\hat{\theta}^{\tau,p}_i$ which is an indicator that retains only wavelets transform which ``strength" is not consistent with a white noise process. In the wavelet setting, ``strength" is defined by the power spectrum. The latter is defined at a particular time and period as the square of the local amplitude of the corresponding wavelet transform component: $P_i(\tau,p)=p|Wave_i(\tau,p)|^2$, where $|\cdot|$ stands for the modulus. 
It is important to note that if no relevant oscillations are detected for some subject at a duration interval of interest, $\hat{s}^{p_1-p_2}_i(\tau)\equiv0$. \\ The duration windows of interest for potassium chosen were 2–14 days, 15–30 days, 31–90 days, 91–180 days and 181–365 days. The rationale behind the choice of the intervals was the clinical interpretation of the short-term components. These intervals were discussed with the clinical experts involved in the study, and they reflect clinical practice routines. More specifically, potassium measurements are usually repeated within 14 days and again after 30 days during up-titration of the pharmacological treatments or more frequently when patients are hospitalized. In both cases, these can be considered possible reasons for the ``acute crisis" in the values of potassium. On the other hand, 90, 180 and 365 days are the intervals where patients can be reevaluated by cardiologists and when effects of disease progression or worsening of comorbidities (e.g. chronic kidney disease) on potassium can be observed. 

Now in the landmark model, we assume that the risk of death does not only depend on the value of the mean subject-specific trajectory of the biomarker estimated through the linear mixed model as in the landmark mixed model in Table \ref{landmixed}  but also on the short-term changes estimated at landmark times using the Morlet wavelet filter described above. Specifically, the landmark wavelet Cox model is now specified as:

\begin{equation}
 h_i(l|\textbf{z}_i,y_{i0},...,y_{ih} , h,w)=h_{h,0}(l) \exp\{\textbf{z}_i\bbeta+\hat{m}_{ih}\gamma_1 + f(\hat{\textbf{s}}_{il})\btheta \}, h\le l\le h+w
\end{equation}

where $\textbf{s}_{il}$ is the vector of local changes in the biomarker at different duration intervals of interest for the subject $i$.  In this case, two alternative models were considered: one with the cubic-spline transformation function as $f(\cdot)$, the other with a categorical version  defined as follows:

\begin{equation*}
\centering
    \hat{o}_i(h)^{p_1-p_2}=
    \begin{cases}
     ``upward" &  \text{if $\hat{s}_i(h)^{p_1-p_2} > 0$ } \\
    ``downward" &  \text{if $\hat{s}_i(h)^{p_1-p_2} < 0$ } \\
    ``no" & \text{if $\hat{s}_i(h)^{p_1-p_2} = 0$} \\
    \end{cases}
\end{equation*}

As in the previous landmark models, the dynamic prediction follows for subject $k$ from the fitted Cox model as:

\begin{equation}
    \hat{\pi}_k^{WAVE}(w|t)=exp[-\hat{H}_{t,0}(w)\exp\{\textbf{z}_k\hat{\bbeta}+\hat{m}_{kt}\hat{\gamma}_1 + f(\hat{\textbf{s}}_{kt})\hat{\btheta}\}]
\end{equation}

\section{Simulation Study}

In the previous section, we discussed dynamic survival models that leverage information from repeated measures of potassium. However, the effectiveness of these models heavily relies on the chosen method employed to obtain the individuals' biomarker trajectories.  We propose the utilization of the wavelet filter, which is commonly employed in the medical signal and time-series domain. By combining the well-established linear mixed model with the wavelet filter, we propose a novel methodology for representing biomarker data with irregular intervals between measurements. 

To assess the performance of this approach, we conducted a simulation study with two main objectives. Firstly, we aimed to compare different specifications of the linear mixed model, with a particular focus on the degrees of freedom of the spline transformation for time. This analysis allowed us to investigate the impact of various model configurations on the accuracy and flexibility of capturing the underlying patterns in the potassium trajectories.
Secondly, we sought to evaluate the effectiveness of the Wavelet Morlet filter adapted for biomarker data obtained from electronic health records, as discussed in the previous section. By incorporating this filter into the modelling process, we aimed to exploit its ability to handle irregular intervals and capture both high-frequency and low-frequency components of the potassium trajectory. 

For the simulation study, we generated 100 datasets, each containing the true potassium trajectories of 1000 individuals. To ensure the synthetic data closely resembled the observed data in terms of heterogeneity and irregularities, we sampled the observations based on the empirical distribution of the measurements found in the actual dataset. Additionally, the maximum follow-up time for each individual was sampled from a Weibull distribution with a scale parameter equal to 3083 and a shape parameter of 1.80.

Following the fitting of linear mixed models and applying the Wavelet filter, we obtained predicted values for the potassium trajectories based on each model. To quantify the accuracy of the predictions, we calculated the Mean Squared Error (MSE) and compared it across the different methods utilized.

\begin{figure}[!h]
    \centering\includegraphics[width=13cm]{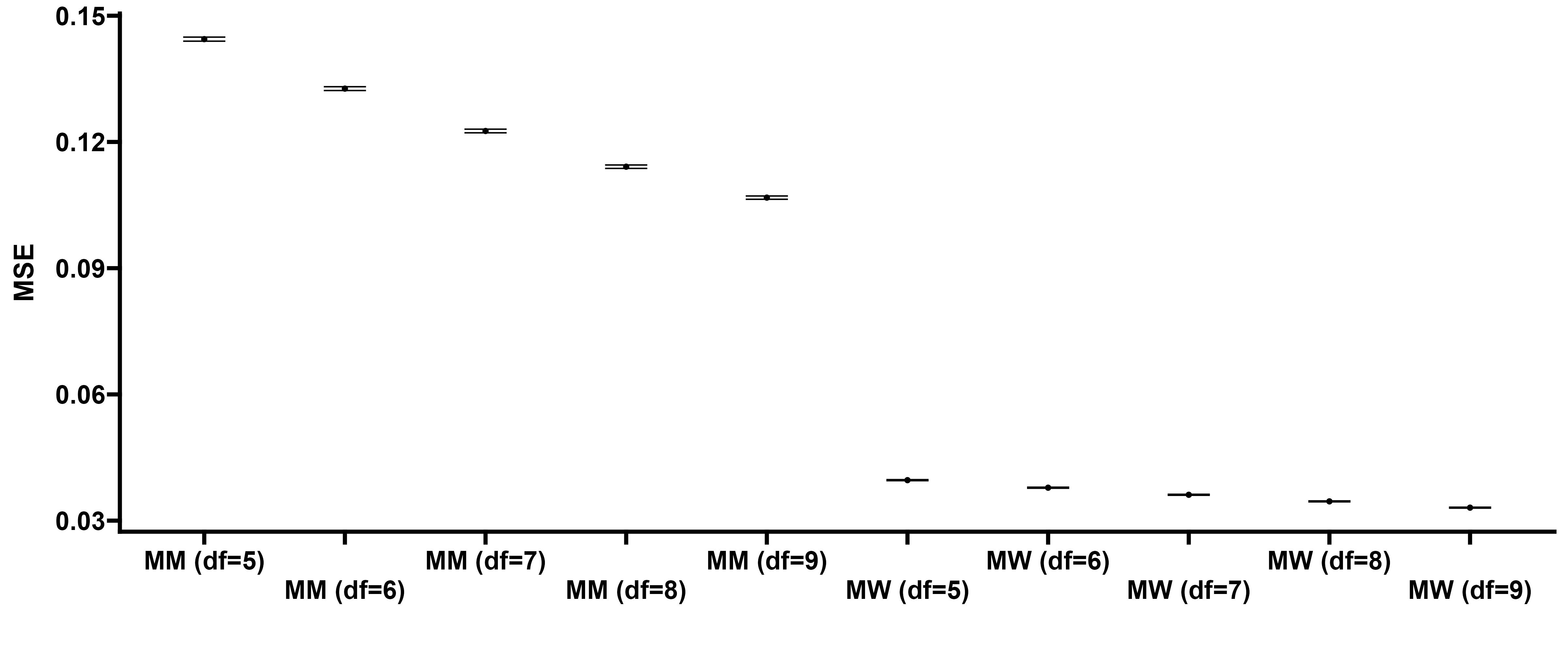}
    \caption{Mean Squared Error (MSE) obtained in the simulation study for the Mixed Model (MM), and the Morlet Wavelet-based method (MW).}
    \label{fig0}
\end{figure}

The results are displayed in Fig. \ref{fig0}. Overall, the proposed approach based on the Wavelet Morlet (MW) filter showed the highest performance in capturing the true biomarker individual profiles compared to the linear Mixed Model (MM). Importantly, contrary to the Mixed Model approach, MW is robust to the choice of the number of degrees of freedom used for the splines. 

\section{Application}

\label{sec4}

\subsection{Study Cohort}

2981 subjects were included in the analysis. Their characteristics at the index date are summarized in Table \ref{table1}.  The median age in the cohort was 77 years (IQR: 70-83) and 58\% were male. The median Left Ventricular Ejection Fraction (LVEF) was 54\% (IQR: 40-63). LVEF is a measure of heart function, and it is considered to be normal when it is over 50\%. In this cohort, 40\% of patients had a reduced LVEF. The New York Heart Association (N.Y.H.A) class indicates the progression of heart failure. In this case, 88\% of patients had a class of 1 or 2 corresponding to asymptomatic and mild symptoms respectively. Furthermore, 48\% of patients had a diagnosis of Chronic Kidney Disease (CKD) and 64\% of patients had more than 3 non-cardiac comorbidities. The median time of the follow-up period in the cohort was 55 months (IQR: 33-80 months). 
Finally, subjects had a median number of potassium measurements of 20 (IQR 11-36) at a median distance of 4 days (IQR: 1-32 days). At 2 years of follow-up, Kaplan-Meier estimate of the survival probability was equal to 0.92 (95\%CI 0.91-9.93).

\begin{table}[!h]
\centering
\caption{Descriptive statistics of the cohort.\label{table1}}
\begin{tabular}{cc}
\textbf{Variable} & \textbf{n = 2,981} \\
\hline
  Age, years & 77 (70, 83) \\ 
  Gender, male & 1,730 (58\%) \\ 
  LVEF, \% & 54 (40, 63) \\ 
  LVEF class &  \\ 
  HFpEF & 1,780 (60\%) \\ 
  HFrEF & 1,201 (40\%) \\
  NYHA Class &  \\ 
  1 & 1,102 (37\%) \\ 
  2 & 1,516 (51\%) \\ 
  3 & 337 (11\%) \\ 
  4 & 26 (0.9\%) \\ 
  CKD & 1,422 (48\%) \\ 
  Non-cardiac comorbidities $>$ 3 & 1,915 (64\%) \\ 
  Number of potassium measurements & 20 (11, 36) \\
  \hline
  Median (IQR); n (\%) & \\
\end{tabular}
\end{table}

\subsection{Longitudinal potassium trajectories}

\begin{figure}[!h]
    \centering\includegraphics[width=10cm]{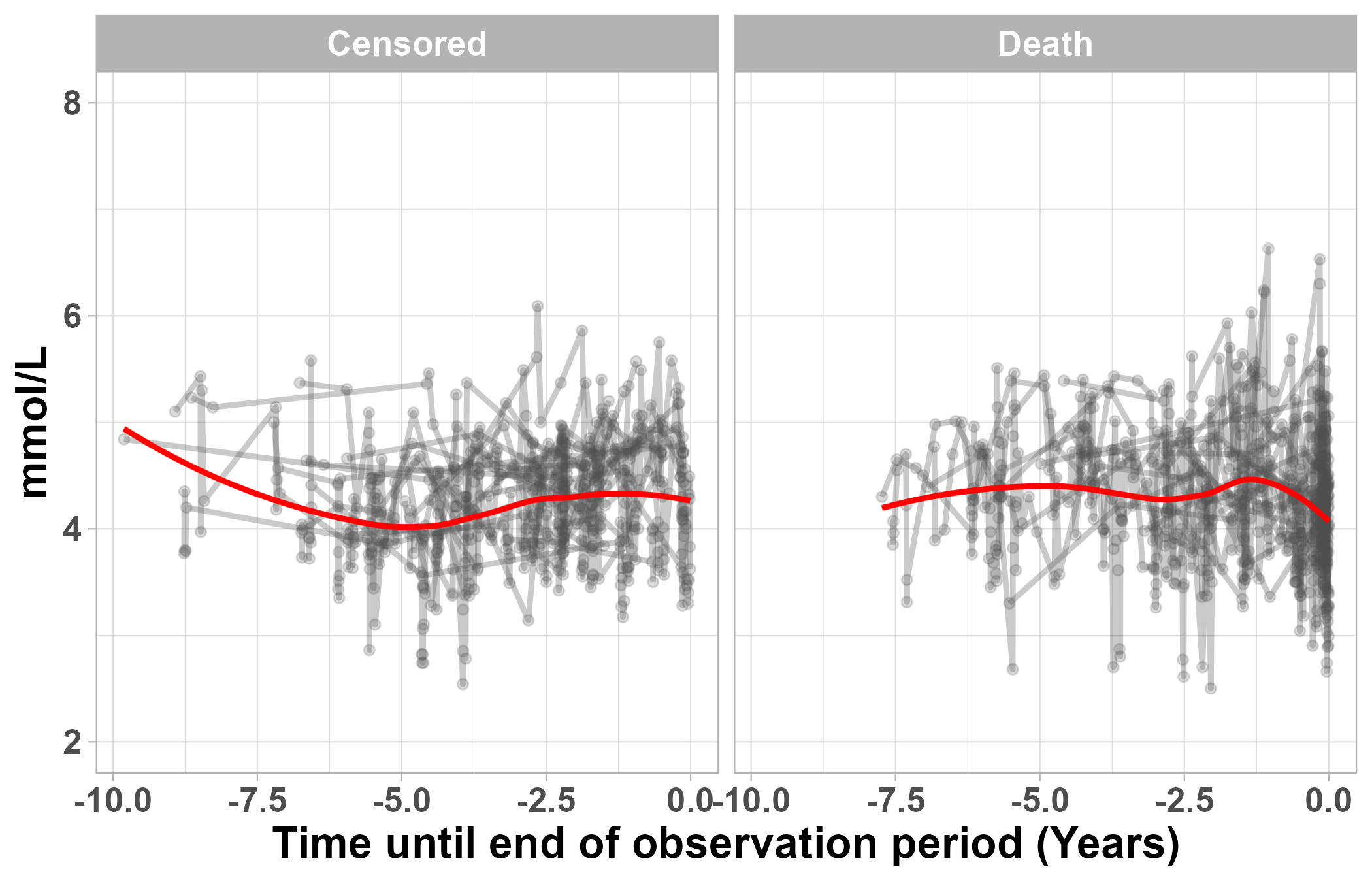}
    \caption{Potassium measurements in a random sample of 50 patients in reverse time according to whether patients died (right panel) or they were censored (left panel). The red lines indicate the lowess smoothing curve. }
    \label{fig1}
\end{figure}

To explore the relationship between the risk of death and potassium (K) in Fig. \ref{fig1} we show the trajectories of potassium observed for a random sample of 50 patients against ``time-to-the-end-of-the-observation-period". Patients have been divided into two groups: subjects who died and patients still alive at the end of the study. A lowess smoothing curve was used to compare the overall trajectory between the two groups. It can be observed that the lowess curve falls within the standard normality range of 3.5-5 mmol/L in both groups. Moreover, the lowess curves appear rather flat over time both in censored and in subjects who died, even though the group with the event shows a slight decrease in potassium values just before the time of death. From the individual observed trajectories, it can be also observed how patients who died tend to exhibit an oscillatory pattern towards the end of the follow-up. 

\subsection{Dynamic survival models}

In all the landmark models presented in the following, as landmark points, we considered time points from 1 to 5 years at a distance of  7  days. The time horizon was fixed at 6 months. As fixed covariates, relevant prognostic factors in heart failure we considered: age, sex, N.Y.H.A class, LVEF class, diagnoses of CKD, and a burden of non-cardiac comorbidities greater than three. For abbreviations, for potassium, the same notation used in the models presented in the Methods Section will be used in the tables. Moreover, $cs(\cdot)$ will denote that a cubic B-spline was used as transformation $f(\cdot)$.

\subsubsection{Landmark LOCF}

\begin{table}[!h]
\caption{Estimates of LOCF1 landmark model coefficients along with standard errors, 95\% confidence intervals and,  p-value.}
\centering
\begin{tabular}{lcccc}
  \hline
\textbf{Variable} & \textbf{Coefficient} & \textbf{Std.Error} & \textbf{95\% CI} & \textbf{p-value} \\
  \hline
Age & 0.67 & 0.05 & 0.57;0.77 & $<$0.01 \\ 
  Sex, Male & 0.30 & 0.07 & 0.16;0.44 & $<$0.01 \\ 
  NYHA III-IV (vs. I-II) & 0.62 & 0.09 & 0.44;0.8 & $<$0.01 \\ 
  HFrEF & 0.12 & 0.07 & -0.02;0.26 & 0.09 \\ 
  $>$3 comorbidities & 0.41 & 0.08 & 0.26;0.57 & $<$0.01 \\ 
  $cs(y_{ih})$ & -1.37 & 0.25 & -1.85;-0.89 & $<$0.01 \\ 
   \hline
\end{tabular}
\label{table2}
\end{table}

\begin{figure}[!h]
    \centering\includegraphics[width=10cm]{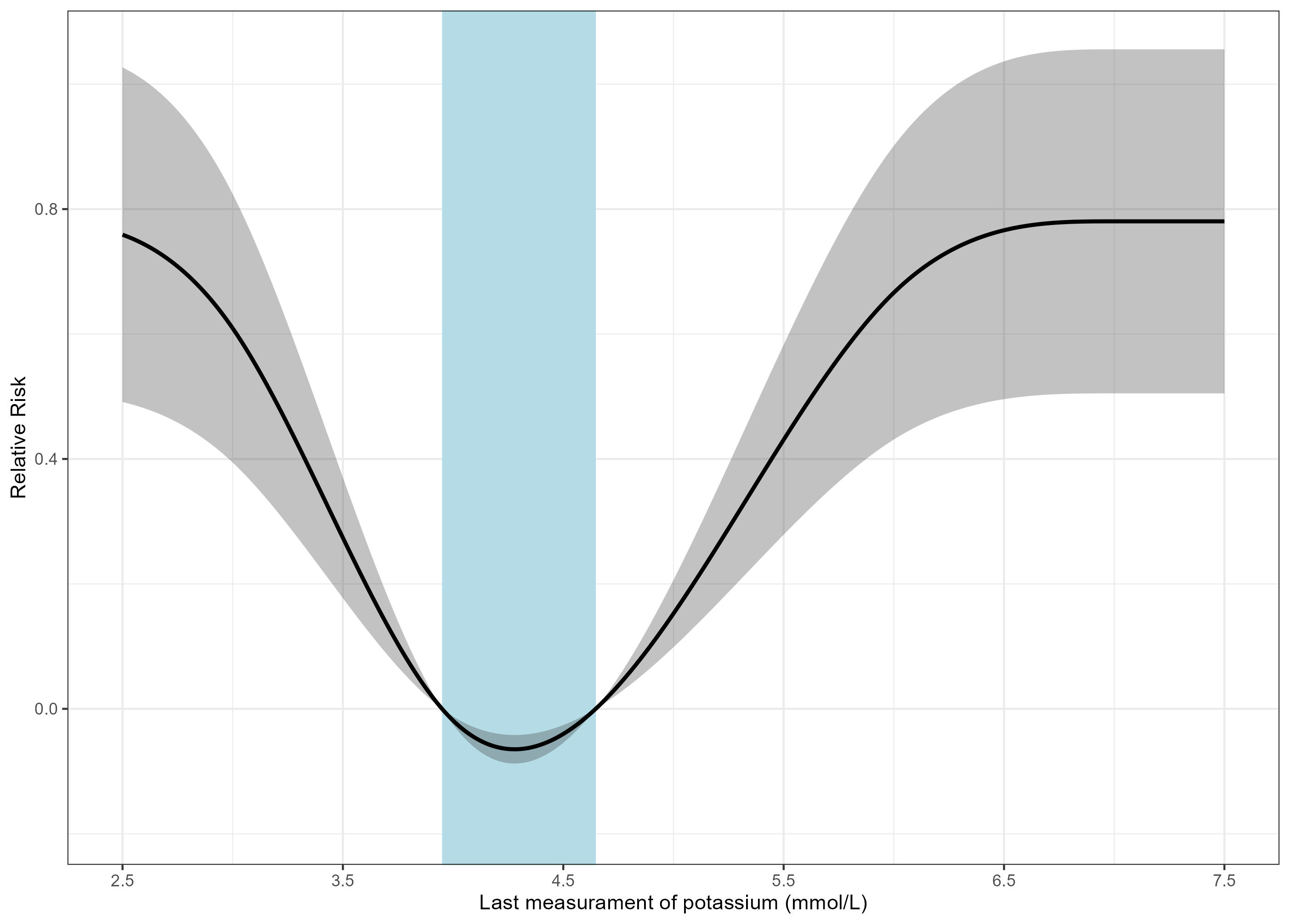}
    \caption{Relative risk of the last value of potassium on the risk of death. }
    \label{fig2}
\end{figure}

The models were fitted using the \texttt{R} \texttt{survival} package \cite{survival-book,survival-package}. First, the two different landmark models, LOCF1 and LOCF2 were fitted to the data. 
Results for the model, with the continuous version of the last observation of potassium before the landmark point, are reported in  Table \ref{table2}. Coherently with previous literature, all fixed covariates are risk factors, increasing the risk of death in heart failure patients. The last value of potassium as a continuous variable is also significant. Because of the spline transformation, its effect on the risk of death is shown in Fig.\ref{fig2}. As expected, both high value and low value of potassium are associated with a greater hazard of dying in the next 6 months. Interestingly, the ``safer" range, shown in blue, is even smaller than current normality cut-offs.

According to the results of the model LOCF2 (Table \ref{table3}), having the last measurement of potassium below 3.5 mmol/L is associated with an increased risk of death, while having the last measurement of potassium greater than 5 mmol/L is weakly associated with a higher hazard of death.

\begin{table}[!h]
\caption{Estimates of LOCF2 landmark model coefficient along with standard errors, 95\% confidence intervals and,  p-value.}
\centering
\begin{tabular}{lcccc}
  \hline
\textbf{Variable} & \textbf{Coefficient} & \textbf{Std.Error} & \textbf{95\% CI} & \textbf{p-value} \\
  \hline
Age & 0.68 & 0.05 & 0.58;0.78 & $<$0.01 \\ 
  Sex, Male & 0.31 & 0.07 & 0.17;0.45 & $<$0.01 \\ 
  NYHA III-IV (vs. I-II) & 0.60 & 0.09 & 0.42;0.77 & $<$0.01 \\ 
  HFrEF & 0.14 & 0.07 & 0;0.29 & 0.05 \\ 
  $>$3 comorbidities & 0.44 & 0.08 & 0.28;0.59 & $<$0.01 \\ 
  $y_{ih}>5$ mmol/L & -0.02 & 0.09 & -0.19;0.15 & 0.83 \\ 
  $y_{ih}<3.5$ mmol/L & 0.89 & 0.12 & 0.65;1.12 & $<$0.01 \\ 
   \hline
\end{tabular}
\label{table3}
\end{table}

\subsubsection{Mixed Landmark}

\begin{figure}[!h]
    \centering\includegraphics[width=10cm]{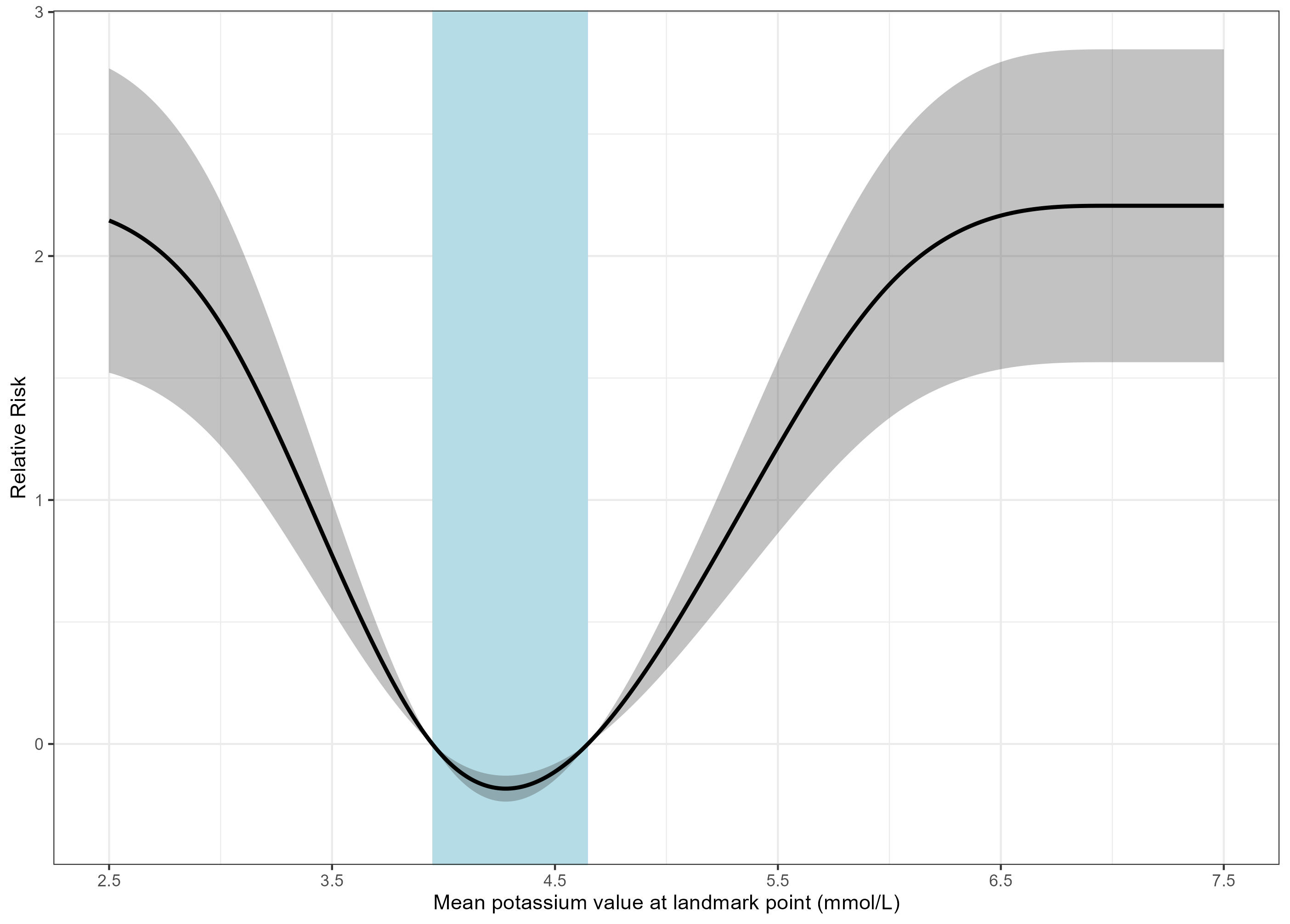}
    \caption{Relative risk of the mean potassium value at the landmark point. }
    \label{fig4}
\end{figure}

The factors which are known to influence potassium are age, sex, Chronic Kidney Disease and heart failure progression. For this reason, these variables, measured at baseline, were used as covariates in the linear mixed model for potassium. The estimates of the fixed coefficients show that males have a higher mean potassium value, C.K.D  increases the mean potassium values, whereas an N.Y.H.A. class above 2 is associated with lower potassium values. The degrees of freedom for the spline transform of time were selected using the AIC. The model with a spline with 9 degrees of freedom was selected.  The complete list of the estimates of the linear effect model is reported in the Appendix.
The results of the mixed landmark model are reported in \ref{table5}. The estimates for the fixed covariates are consistent with the results of the previous model. Moreover, as shown in Fig.\ref{fig4}, the relative risk has an u-shape relationship with different values of the individual mean potassium trajectory.

\begin{table}[ht]
\caption{Estimates of mixed landmark model coefficients along with standard errors, 95\% confidence intervals and,  p-value.}
\centering
\begin{tabular}{lrrll}
  \hline
\textbf{Variable} & \textbf{Coefficient} & \textbf{Std.Error} & \textbf{95\% CI} & \textbf{p-value} \\
  \hline
Age & 0.68 & 0.05 & 0.58;0.77 & $<$0.01 \\ 
  Sex, Male & 0.30 & 0.07 & 0.16;0.44 & $<$0.01 \\ 
  NYHA III-IV (vs. I-II) & 0.62 & 0.09 & 0.44;0.79 & $<$0.01 \\ 
  HFrEF & 0.13 & 0.07 & -0.02;0.27 & 0.08 \\ 
  $>$3 comorbidities & 0.39 & 0.08 & 0.23;0.54 & $<$0.01 \\ 
  $cs(\hat{m_{ih}})$ & -3.87 & 0.57 & -5;-2.75 & $<$0.01 \\
   \hline
\end{tabular}
\label{table5}
\end{table}

 \subsubsection{Joint Model}
 
\begin{figure}[!h]
    \centering\includegraphics[width=10cm]{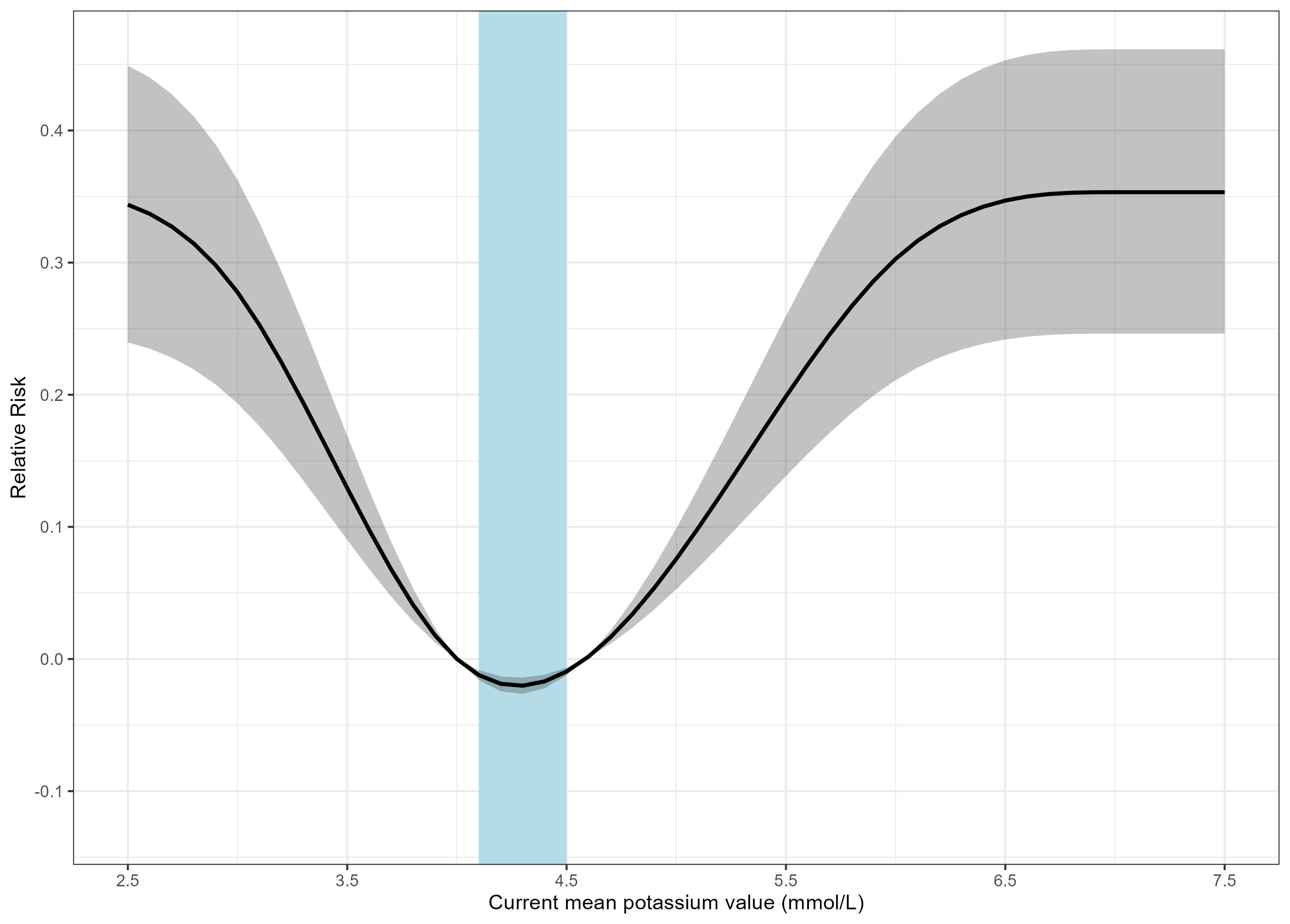}
    \caption{Relative risk of current mean value of potassium}
    \label{fig5}
\end{figure}

To model the survival and the potassium process jointly, the joint model has been fitted with the package \texttt{JMBayes2} \cite{JMbayes2}. For the longitudinal part of the linear mixed effect model, the same specification of the mixed-landmark model was used, with the exception that four degrees of freedom were used for the splines. Due to the computational complexity of the joint model, it was not possible to consider a higher number of degrees of freedom.  The results of the longitudinal and the survival parts of the model are reported in Table \ref{table6} and Table \ref{table7} respectively. As can be seen, the association between potassium and the risk of death is quite similar to the ones obtained with the mixed landmark model of the previous section (Fig. \ref{fig5}).

\begin{table}[!h]
\caption{Estimates of longitudinal part of the joint model coefficient along with standard errors, 95\% confidence intervals and,  p-value.
 }
\centering
\begin{tabular}{lcccc}
  \hline
\textbf{Variable} & \textbf{Coefficient} & \textbf{Std.Error} & \textbf{95\%CI} & \textbf{p-value} \\ 
  \hline
  Sex, Male & 0.06 & 0.01 & 0.03;0.09 & $<$0.01 \\ 
  Age & -0.01 & 0.01 & -0.02;0.01 & 0.41 \\ 
  CKD & 0.12 & 0.01 & 0.1;0.15 & $<$0.01 \\ 
  NYHA III-IV (vs. I-II) & -0.06 & 0.02 & -0.1;-0.02 & $<$0.01 \\ 
  $ns_1(t)$ & 0.03 & 0.04 & -0.06;0.11 & 0.54 \\ 
 $ns_2(t)$ & -0.02 & 0.09 & -0.21;0.16 & 0.8 \\ 
  $ns_3(t)$ & 0.08 & 0.12 & -0.16;0.32 & 0.51 \\ 
  $ns_4(t)$ & -0.18 & 0.27 & -0.71;0.35 & 0.49 \\ 
   \hline
\end{tabular}
\label{table6}
\end{table}

\begin{table}[!h]
\centering
\caption{Estimates of survival part of the joint model coefficients along with standard errors, 95\% confidence intervals and,  p-value.}
\begin{tabular}{lcccc}
  \hline
\textbf{Variable} & \textbf{Coefficient} & \textbf{Std.Error} & \textbf{95\% CI} & \textbf{p-value} \\
  \hline
  Age & 0.69 & 0.07 & 0.56;0.82 & $<$0.01 \\ 
  Sex, Male & 0.34 & 0.06 & 0.21;0.47 & $<$0.01 \\ 
  NYHA III-IV (vs. I-II) & 0.44 & 0.08 & 0.28;0.58 & $<$0.01 \\ 
  HFrEF & 0.09 & 0.06 & -0.03;0.21 & 0.13 \\ 
  $>$3 comorbidities & 0.40 & 0.06 & 0.28;0.53 & $<$0.01 \\ 
  $cs(m_{it})$ & -0.61 & 0.09 & -0.79;-0.42 & $<$0.01 \\ 
  
   \hline
\end{tabular}
\label{table7}
\end{table}

\subsubsection{Landmark mixed-wavelet}

\begin{figure}[!h]
    \centering\includegraphics[width=13cm]{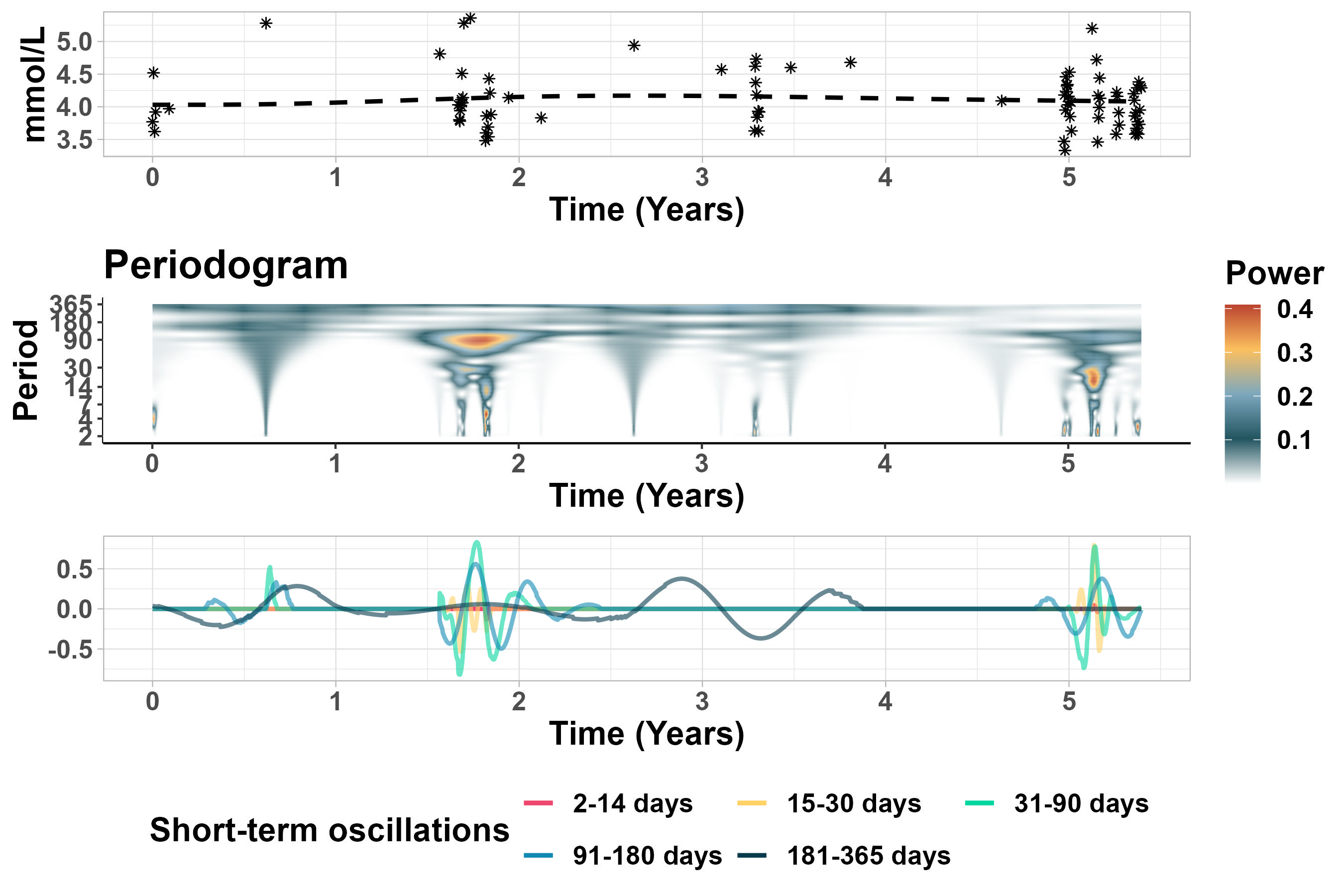}
    \caption{\textbf{Top panel}: Potassium measurements over time and mean current value (dashed line) for one patient. \textbf{Central panel}: Periodogram obtained through the wavelet transform. \textbf{Bottom panel}: Example of corresponding short-term oscillations obtained through wavelet filtering.}
    \label{fig6}
\end{figure}

 Following the steps described in Section 3.5, we obtained the vector of the continuous and categorical version of short-term changes of potassium at the landmark points.   As an example, the periodogram and the derived short-term oscillations together with the potassium measurements and the mean current value for one subject are presented in Fig.\ref{fig6}.
 
Over the 5 years of observations, the mean trajectory estimated using the linear mixed effect model (top panel, dashed line) is rather flat. However, there are some sudden changes that are captured by the periodogram (central panel). In correspondence to the stronger power spectrum areas, the filtering captures some changes in the interval 181–365 days and between 15 and 90 days.

Both the landmark model with the continuous covariates representing the short-term changes and the one with the corresponding categorical ones were fitted. The one with the categorical variables had the lowest AIC and therefore its results are reported in Table \ref{table8}.

\begin{table}[!h]
\centering
\caption{Estimates of mixed-wavelet landmark model coefficients along with standard errors, 95\% confidence intervals and,  p-value.}
\begin{tabular}{lcccc}
  \hline
\textbf{Variable} & \textbf{Coefficient} & \textbf{Std.Error} & \textbf{95\% CI} & \textbf{p-value} \\
  \hline
Age & 0.67 & 0.05 & 0.57;0.77 & $<$0.01 \\ 
  Sex, Male & 0.28 & 0.07 & 0.14;0.42 & $<$0.01 \\ 
  NYHA III-IV (vs. I-II) & 0.61 & 0.09 & 0.44;0.79 & $<$0.01 \\ 
  HFrEF & 0.15 & 0.07 & 0.01;0.29 & 0.03 \\ 
  $>$3 comorbidities & 0.36 & 0.08 & 0.2;0.51 & $<$0.01 \\ 
  $cs(\hat{m}_{ih})$  & -2.79 & 0.59 & -3.96;-1.62 & $<$0.01 \\ 
  $\hat{o}_{ih}^{[2-14]}$: upward & 0.48 & 0.06 & 0.36;0.59 & $<$0.01 \\ 
   $\hat{o}_{ih}^{[2-14]}$: downward & 0.43 & 0.05 & 0.33;0.54 & $<$0.01 \\
   $\hat{o}_{ih}^{[15-30]}$: upward & 0.49 & 0.04 & 0.41;0.58 & $<$0.01 \\ 
   $\hat{o}_{ih}^{[15-30]}$: downward & 0.50 & 0.04 & 0.42;0.59 & $<$0.01 \\ 
  $\hat{o}_{ih}^{[31-90]}$: upward & 0.45 & 0.05 & 0.34;0.55 & $<$0.01 \\ 
  $\hat{o}_{ih}^{[31-90]}$: downward & 0.48 & 0.05 & 0.38;0.58 & $<$0.01 \\ 
  $\hat{o}_{ih}^{[91-180]}$:upward & 0.25 & 0.06 & 0.13;0.37 & $<$0.01 \\ 
  $\hat{o}_{ih}^{[91-180]}$:downward & 0.30 & 0.06 & 0.18;0.42 & $<$0.01 \\ 
  $\hat{o}_{ih}^{[181-365]}$:upward & 0.08 & 0.07 & -0.05;0.21 & 0.22 \\
  $\hat{o}_{ih}^{[181-365]}$:downward & 0.16 & 0.07 & 0.03;0.29 & 0.02 \\
   \hline
\end{tabular}
\label{table8}
\end{table}

As in the previous model, the mean value of potassium estimated through the linear mixed model appears to be significantly associated with the risk of death. Moreover, the presence of both downward and upward past variations in the last year has an impact on the risk of death. Therefore, the history of the longitudinal process seems to influence the survival probability up to 1 year. Both downward and upward oscillations increase the risk of death. Furthermore, those with a duration of less than 3 months have the strongest effect. 

\subsection{Models comparison}

First, the overall goodness-of-fit of the different landmark survival models was compared in terms of AIC, BIC, and C-concordance index obtained with 10-fold Cross-Validation (Table \ref{table10}). The landmark mixed-wavelet model shows the lowest AIC and BIC. Moreover, the C-index obtained with this approach is significantly higher than the others.

\begin{table}[ht]
\centering
\begin{tabular}{lccc}
  \hline
\textbf{Model} & \textbf{AIC} & \textbf{BIC} & \textbf{C-Index (95\% CI}) \\ 
  \hline
Landmark LOCF1 & 315900& 315948 & 0.692 (0.674-0.711) \\ 
  Landmark LOCF2 & 31576 & 315825& 0.691 (0.674-0.709) \\ 
  Landmark Mixed & 315232 & 315280 & 0.698 (0.679-0.717) \\ 
  Landmark Wavelet & 310636 & 310763& 0.745 (0.729-0.761) \\ 
   \hline
\end{tabular}
\caption{Comparison of goodness-of-fit between the different landmark models in terms of AIC, BIC and C-index (obtained with 10-fold cross-validation).}
\label{table10}
\end{table}

Then, a comparison of all the dynamic survival models in terms of the dynamic predictive performance was obtained through discrimination and calibration measures. Specifically, we used the Dynamic Area Under the Curve (AUC(t)) and the Brier Score. 10-fold cross-validation was used to obtain both indices. The prediction was made at three different times: 1 year, 2 years and 3 years of follow-up and the prediction horizon considered was 6 months. For the definition of AUC(t) and the Brier Score, the one proposed by \citet{Blanche2015} was used. The results are presented in Table \ref{table9}. It can be observed that the novel wavelet method showed the highest discrimination.

\begin{table}[ht]
\caption{Comparison of predictive accuracy measures between the different dynamic survival models in terms of discrimination and calibration (obtained with 10-fold cross-validation).}
\centering
\begin{tabular}{cccc}
  \hline
    & \textbf{Prediction time: 1 year} &\\
  \hline
 \textbf{Model} & \textbf{AUC (6 months)} & \textbf{}Brier Score (6 months)  \\ 
  \hline
  
    Landmark LOCF1 & 0.51 (0.43-0.59) & 0.039 (0.032-0.045) \\ 
  Landmark LOCF2 & 0.51 (0.47-0.55) & 0.038 (0.032-0.045) \\  
  Landmark Mixed &  0.54 (0.49-0.6) & 0.039 (0.032-0.046) \\ 
  Joint Model & 0.5 (0.44-0.56) & 0.04 (0.033-0.047) \\ 
   Landmark Wavelet & 0.69 (0.64-0.73) & 0.037 (0.031-0.044) \\ 
    \hline
     & \textbf{Prediction time: 2 years} &\\
   \hline
\textbf{Model} & \textbf{AUC (6 months)} & \textbf{}Brier Score (6 months)  \\ 
  \hline
Landmark LOCF1  & 0.59 (0.53-0.65) & 0.041 (0.033-0.049) \\ 
 Landmark LOCF2 & 0.53 (0.47-0.59) & 0.041 (0.033-0.049) \\ 
  Landmark Mixed & 0.6 (0.54-0.66) & 0.041 (0.033-0.049) \\ 
  Joint Model & 0.52 (0.46-0.58) & 0.044 (0.036-0.051) \\ 
  Landmark Wavelet  &  0.72 (0.68-0.77) & 0.039 (0.032-0.047) \\ 
    \hline
     & \textbf{Prediction time: 3 years} &\\
  \hline
\textbf{Model} & \textbf{AUC (6 months)} & \textbf{}Brier Score (6 months)  \\ 
  \hline
   Landmark LOCF1 &  0.51 (0.44-0.57) & 0.051 (0.044-0.057) \\  
   Landmark LOCF2 &   0.51 (0.5-0.52) & 0.051 (0.044-0.058) \\ 
 Landmark Mixed &  0.53 (0.46-0.6) & 0.05 (0.044-0.057) \\ 
  Landmark Wavelet & 0.66 (0.6-0.71) & 0.05 (0.043-0.056) \\ 
  Joint Model & 0.53 (0.46-0.6) & 0.052 (0.045-0.059) \\ 
   \hline
\end{tabular}
\label{table9}
\end{table}

\subsection{Dynamic predictions}

\begin{figure}[!h]
    \centering\includegraphics[width=15cm]{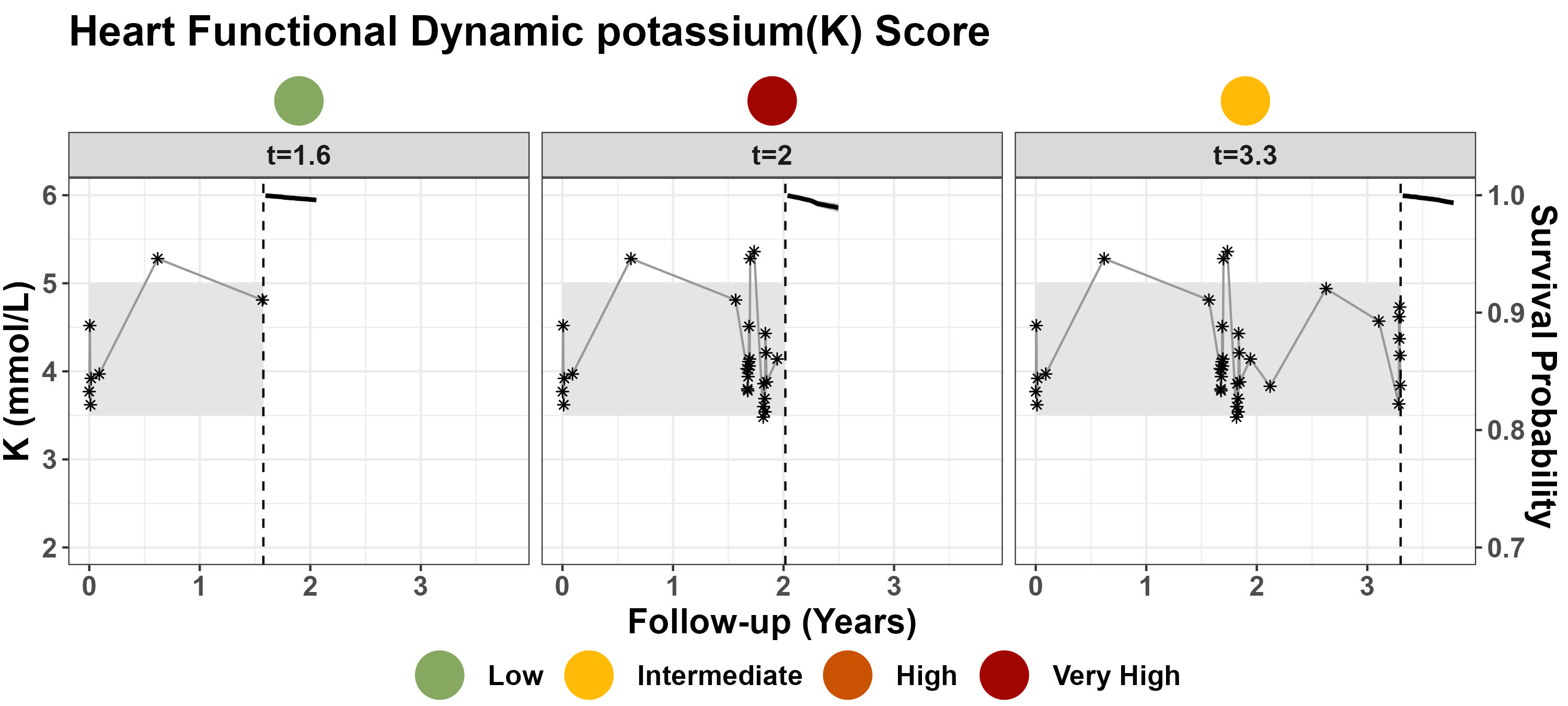}
    \caption{Example of dynamic prediction for one patient's HFDKS and corresponding 6-months survival probability.}
    \label{fig7}
    \end{figure}

In Heart Failure, the monitoring of potassium plays an important role. Therefore, a score which could easily be used by a cardiologist would be very useful to quantitatively assess the patient's risk using the potassium measurements available. Fig.\ref{fig7} shows an example of a patient's predicted survival probability dynamically updated at times 1.6, 2 and 3.3 years. In addition to the dynamic prediction of the survival probability,  we also propose a score based on the wavelet landmark model, which feeds a dynamic tool for supporting medical doctor's daily practice.  To obtain the score, the partial linear predictor, containing all the terms related to potassium, was predicted for an individual $i$ at time $t$ and it was categorized in risk groups by using the observed quartiles in the entire cohort.  We call this the Heart Functional Dynamic potassium (K) Score. Fig.\ref{fig7} also displays the corresponding predicted HFDKS. It can be observed how at one year and a half of follow-up, while having a previous potassium value over 5 mmol/L, the patient is at low risk. The risk becomes very high at two years after the subject has experienced multiple abrupt changes. At time=3.3 years of follow-up, the score decreases as changes have been less extreme. As expected, the predicted 6-month survival probability shows a similar pattern, with a lower probability of survival in the following 6 months at 2 and 3.3 years of follow-up. 

\section{Discussion \& Conclusions}
\label{sec5}

In this paper, we tackled the problem of providing a suitable dynamic survival model to assist in the monitoring of potassium in heart failure patients. In this disease, it is very important to monitor potassium since it is often altered by the disease as well as pharmacological treatments. The relationship between potassium and survival in heart failure remains debated, and the role of its dynamics over time has not been studied yet. Currently, in the clinical practice normality cut-offs on a single measurement are the only available quantitative method to assess patients' situation with regard to potassium. The extensive research carried out in the field of landmark models and joint models has shown the importance of exploiting the information provided by repeated measurements to take into account measurement error and past history of the biomarker process. Existing dynamic survival models have focused on mixed-effects models. However, when the subjects' observation period time spans over years and the biomarker, in this case potassium, has important, but not lasting changes, the linear mixed model is not flexible enough to capture this local variation. Hence, short-term changes are essentially lost in the error term.  Existing literature has not considered that the timescale of the survival process is very different from the timescale at which changes in the biomarkers happen in the human body. Especially in real-world data coming from electronic health records, the follow-up period can be very long (e.g. years), whereas intervals of measurement times are highly irregular because they depend on the clinical requirements. For example, during a drug's up-tritation, more frequent measurements are needed. As a consequence, such data are characterized by a  very high observation period/time-between-measurements ratio. This issue can't be solved by simply considering a more flexible specification of the mixed-effects model because the short-term oscillations do have not a common structure across subjects, instead, they vary across individuals both in terms of timing and duration. \\ 
To overcome this problem, we have proposed a novel method which combines linear mixed-effect modelling coupled with a functional approach based on wavelet filters. The linear mixed model allows capturing the effect of baseline covariates and the subject-specific long-term profiles, while the functional approach on the residuals allows identifying the individual short-term oscillations at different timings and of different durations. The simulation study showed that regardless of the number of degrees of freedom used for the spline transform of time in the linear mixed model, the wavelet approach is more accurate in capturing the true individual potassium profiles based on the available measurements.  \\
In addition, the proposed method enabled us to show the prognostic role of short-term changes in potassium, independently of the effect of the mean value of potassium. Thus, we laid the basis for a dynamic score to quantitatively summarize the risk of death associated with the observed history of potassium measurements. This is an important point since the goal of this work was also to provide a possible solution for the problem of monitoring potassium in clinical practice. In the context of the application presented, we compared the performance in terms of goodness-of-fit and dynamic predictive ability of our method with standard landmarking, a mixed-landmark model and a joint model. From this comparison, it is clear that correctly exploiting information from the longitudinal process is very important. The two methods expected to perform best (the joint model and the mixed landmark survival model) are not better than the LOCF approach, which disregards the past completely. The reason is that the mean individual potassium trajectories are not the only relevant part of the longitudinal process from a prognostic point of view. On the contrary, the wavelet approach recovers the short-term oscillations and improves the performance of the survival model. \\ 
Even though the method is proposed here for a specific clinical application, and we don't claim it to be a general method, we believe that it would be of interest to study it further in other clinical settings involving long-term monitoring of biomarkers. Moments of crises in biomarkers translating into short-term changes can be quite common in chronic diseases, and our approach can help to study their role in a time-to-event setting. Specifically, the method is flexible in the definition of what short-term means. Both the choice of the maximum period to consider for the short-term oscillation and the intervals for the dimensionality reduction of the periodogram with respect to the frequency domain can be chosen according to biological knowledge and sample size considerations. \\
In conclusion, in studying the relationship between potassium and survival, the most important aspect consists in modelling the dependence of the survival process on the biomarker's past. In our work, we were able to explore how much of the history of the potassium process is relevant to the time-to-event process. \\
At the moment, the linear mixed model component and the short-term components are estimated in two steps. In the future, it could be of interest to develop a method to obtain them simultaneously. Moreover, we are aware that the landmark model is a working model, and it would be more satisfactory and coherent from a statistical perspective to introduce a joint model that includes the wavelets in the longitudinal component. However, it would also be far more challenging from a computational point of view, especially given the amount of data present in studies involving electronic health records.  We believe that this work and the promising results obtained in this application with the wavelet landmark prove that it is worth it to keep working in this direction and that there is still space for improvement in the field of dynamic survival models.

\bibliography{refs.bib}

\section*{Acknowledgments}
This work was supported by VIFOR Pharma. The authors also thank Dr.Annamaria Iorio for her cardiological insights.

\section*{Appendix}

\captionsetup[table]{skip=10pt}
\renewcommand{\thetable}{S\arabic{table}}
\setcounter{table}{0}

\begin{table}[ht]
\centering
\caption{Estimates of the linear mixed effect model used for the mixed landmark model.}
\begin{tabular}{lcccc}
  \hline
\textbf{Variable} & \textbf{Coefficient} & \textbf{Std.Error} & \textbf{95\% CI} & \textbf{p-value} \\
  \hline
Intercept & 4.13 & 0.01 & 4.1;4.15 & $<$0.01 \\ 
  Sex, Male & 0.06 & 0.01 & 0.04;0.08 & $<$0.01 \\ 
  Age & -0.00 & 0.01 & -0.02;0.01 & 0.43 \\ 
  CKD & 0.12 & 0.01 & 0.1;0.14 & $<$0.01 \\ 
  NYHA III-IV (vs. I-II) & -0.06 & 0.02 & -0.09;-0.02 & $<$0.01 \\ 
    $ns_1(time)$ & 0.14 & 0.02 & 0.09;0.18 & $<$0.01 \\ 
   $ns_2(time)$& 0.10 & 0.02 & 0.06;0.15 & $<$0.01 \\ 
   $ns_3(time)$ & 0.09 & 0.02 & 0.05;0.13 & $<$0.01 \\ 
    $ns_4(time)$ & 0.06 & 0.02 & 0.02;0.1 & $<$0.01 \\ 
   $ns_5(time)$ & 0.03 & 0.02 & -0.01;0.07 & 0.13 \\ 
  $ns_6(time)$ & 0.00 & 0.04 & -0.08;0.08 & 0.96 \\ 
    $ns_7(time)$ & 0.07 & 0.05 & -0.04;0.17 & 0.2 \\ 
    $ns_8(time)$ & -0.17 & 0.11 & -0.38;0.05 & 0.13 \\ 
   \hline
\end{tabular}
\end{table}

\begin{table}[ht]
\centering
\caption{Estimates of the linear mixed effect model used for the wavelet landmark model.}
\begin{tabular}{lcccc}
  \hline
\textbf{Variable} & \textbf{Coefficient} & \textbf{Std.Error} & \textbf{95\% CI} & \textbf{p-value} \\
  \hline
Intercept & 4.13 & 0.01 & 4.11;4.16 & $<$0.01 \\ 
  Sex, Male & 0.06 & 0.01 & 0.04;0.08 & $<$0.01 \\ 
  Age & -0.01 & 0.01 & -0.02;0.01 & 0.33 \\ 
  CKD & 0.12 & 0.01 & 0.1;0.14 & $<$0.01 \\ 
  NYHA III-IV (vs. I-II) & -0.05 & 0.02 & -0.09;-0.02 & $<$0.01 \\ 
  $ns_1(time)$ & 0.07 & 0.02 & 0.04;0.1 & $<$0.01 \\ 
  $ns_2(time)$ & 0.06 & 0.02 & 0.02;0.09 & $<$0.01 \\ 
 $ns_3(time)$ & -0.07 & 0.03 & -0.14;0 & 0.04 \\ 
  $ns_4(time)$ & 0.13 & 0.04 & 0.05;0.22 & $<$0.01 \\ 
  $ns_5(time)$ & -0.07 & 0.09 & -0.24;0.1 & 0.43 \\ 
   \hline
\end{tabular}
\end{table}

\end{document}